\documentclass[11pt]{article}
\usepackage[preprint]{acl}
\usepackage{times,latexsym}
\usepackage[T1]{fontenc}
\usepackage[utf8]{inputenc}
\usepackage{stfloats}
\usepackage{microtype,inconsolata,graphicx,booktabs,multirow,tabularx,placeins}
\usepackage[table]{xcolor}
\usepackage[breakable,listings]{tcolorbox}

\newtcblisting{promptbox}[2][]{
  listing only,breakable,title={#2},
  colback=black!2,colframe=black!40,
  colbacktitle=black!6,coltitle=black,
  fonttitle={\raggedright\hyphenpenalty=10000\relax
    \bfseries\normalsize},
  boxrule=0.4pt,arc=1mm,
  left=3pt,right=3pt,top=3pt,bottom=3pt,
  listing options={
    basicstyle=\ttfamily\normalsize,
    breaklines=true,
    columns=fullflexible,
    keepspaces=true,
    showstringspaces=false,
    breakatwhitespace=false
  },
  #1
}

\newtcolorbox{questionexample}[2][]{title={#2},
  colback=black!2,colframe=black!40,colbacktitle=black!6,coltitle=black,
  fonttitle={\raggedright\hyphenpenalty=10000\relax\bfseries\normalsize},
  fontupper=\normalsize,
  fontlower=\normalsize,
  before upper=\raggedright,before lower=\raggedright,
  boxrule=0.4pt,arc=1mm,
  left=5pt,right=5pt,top=5pt,bottom=5pt,#1}

\title{Astro2Method: Benchmarking LLMs on Machine Learning Method Selection in Astronomy}
\author{
  Xingzhong Fan \quad Hangfeng He \quad Fatemeh Nargesian \\
  Department of Computer Science, University of Rochester \\
  Rochester, NY, USA \\
  \texttt{xfan17@ur.rochester.edu} \\
  \texttt{\{hangfeng.he, fnargesian\}@rochester.edu}
}
\begin{document}
\maketitle
\begin{abstract}

Machine learning (ML) is increasingly used in astronomy, but choosing a suitable method for a scientific problem often requires expertise in both astronomy and ML. Large language models (LLMs) could help astronomers make such choices, yet this ability remains underexplored in existing benchmarks. We introduce Astro2Method, a benchmark of 500 expert-reviewed binary questions from 346 astronomy papers (2024/01–2026/06), based on clear performance differences and complete (Easy) or partial (Hard) explanations from the authors. Across five LLMs, direct prompting achieves 87.50–92.55\% accuracy on Easy questions but only 58.87–64.52\% on Hard.  Using historical papers, neither retrieval-augmented generation nor most agent methods with additional reasoning or retrieval (including ReAct, Self-Ask, and CRITIC) bring consistent improvement. Agents required to search the web improve accuracy mainly when they find the papers the questions are based on, with no consistent improvement otherwise. Providing the source paper's pipeline description with method names masked improves Hard accuracy, matching or exceeding web search. Together, these results suggest that relevant experimental information helps distinguish ML methods, but inferring it from related studies and domain knowledge remains challenging for LLMs in astronomy. \footnote{The Astro2Method benchmark dataset is publicly available at \url{https://huggingface.co/datasets/Xingzhongfan/Astro2Method}.}

\end{abstract}
\section{Introduction}
\label{sec:introduction}

With the rapid growth of astronomical data, machine learning (ML) is increasingly used in astronomy, contributing to discoveries such as the Galactic neutrino signal observed by IceCube~\cite{ting2025deep,icecube2023observation}. 
However, choosing an ML method often requires expertise in both astronomy and ML. Large language models (LLMs) can assist astronomical research, but they can also make mistakes~\cite{sokol2026last}. Can they reliably help astronomers choose ML methods?

Existing LLM benchmarks in astronomy, such as AstroMLab~\cite{ting2025astromlab} and Astro-QA~\cite{li2025astronomical}, mainly assess LLMs' astronomical knowledge; AstroVisBench~\cite{joseph2026astrovisbench} evaluates code generation for astronomical data processing and visualization. Broader scientific benchmarks evaluate LLMs on competition and college-level questions~\citep{qiu2026phybench,he2024olympiadbench,wang2023scibench}. This leaves a gap in evaluating LLMs' ability to select ML methods.

To address this gap, we introduce Astro2Method, a benchmark of 500 binary questions asking which ML method performs better for a given astronomical problem. Candidate methods may differ in model design, training strategy, or data processing. We screened 1,028 astronomy papers published between 2024/01/01 and 2026/06/30, selecting ML method comparisons from 346 papers based on clear performance differences and complete (Easy) or partial (Hard) explanations from the authors. All questions were reviewed and revised by human experts; 376 are Easy and 124 are Hard.

We test five LLMs with direct prompting and three with retrieval-augmented generation (RAG) and agent methods. Direct prompting shows a large accuracy gap between Easy and Hard. With retrieval limited to historical papers, RAG brings no consistent improvement, while agent methods using ReAct, Self-Ask, and CRITIC score 1.4--6.2 percentage points (pp) below direct prompting in Overall accuracy. In contrast, agents using web search can access the papers underlying the questions, with gains concentrated when those papers are found. Providing only the source paper's pipeline with method names masked improves Hard accuracy by 12.1--14.5 pp over direct prompting. These results suggest that relevant experimental information helps distinguish ML methods, but inferring it from related studies and domain knowledge remains challenging for LLMs in astronomy.

\section{Dataset Construction}
\label{sec:benchmark}

\begin{figure*}[t]
  \centering
  \includegraphics[width=\textwidth]{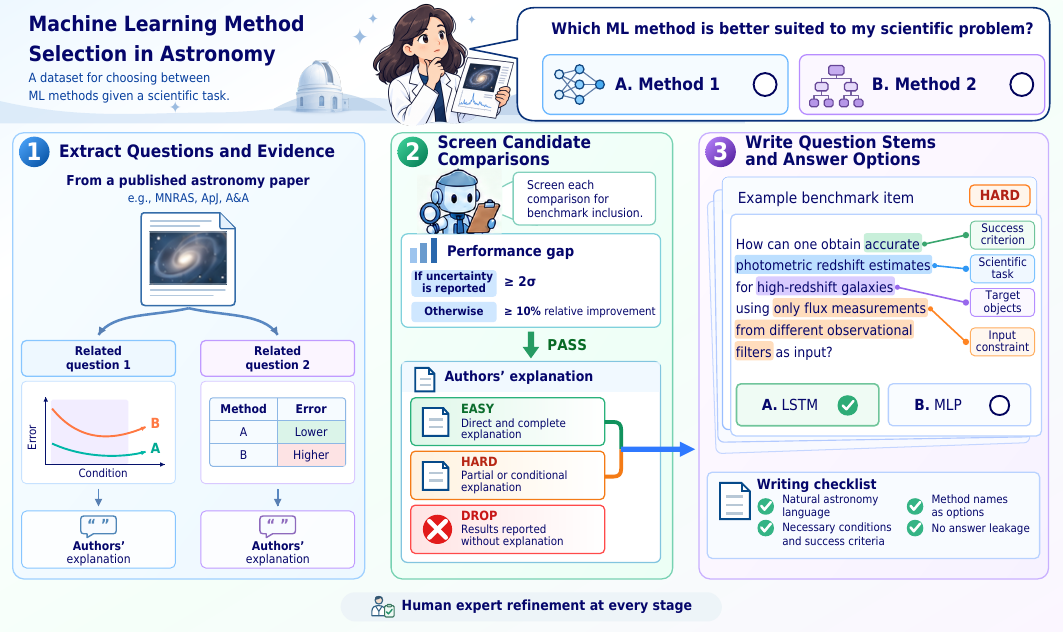}
  \caption{Dataset construction pipeline. We extract comparisons from astronomy papers, screen them using performance differences and authors’ explanations, and write binary questions with expert review throughout.}
  \label{fig:benchmark-construction}
\end{figure*}

Each item in our dataset contains an astronomical question and two candidate ML methods, which may differ in model design, training strategy, or data processing. We searched NASA's Astrophysics Data System
(\href{https://ui.adsabs.harvard.edu/}{ADS}) for papers published from 2024/01/01 to 2026/06/30 in major astronomy journals, using “neural network” in titles, abstracts, or keywords (see Appendix~\ref{app:literature-collection}). 
Through iterative review of full-paper PDFs with LLM assistance, we identify scientific questions, candidate ML method comparisons, and authors' explanations (Appendix~\ref{app:question-construction-interaction}); one paper may contribute multiple questions.
We select candidate method pairs based on performance differences as follows.
If the source paper reports the significance of the performance difference, we require at least two standard deviations.
Otherwise, if standard deviations $\sigma_A$ and $\sigma_B$ are reported for scores $m_A$ and $m_B$, we use standard error propagation and require $|m_A-m_B|\geq 2\sqrt{\sigma_A^2+\sigma_B^2}$.
If neither type of uncertainty information is available, we require a relative improvement of at least $10\%$.
For higher-is-better metrics with a theoretical maximum $U$, we measure relative improvement by the relative reduction in the gap to $U$.

Next, human experts assign difficulty based on the authors' explanations: direct and complete explanations yield Easy questions, partial explanations yield Hard, and comparisons without explanations are dropped. We then turn each comparison into a binary question asking which method better suits the scientific problem and setup. Throughout construction, human experts review and refine candidate comparisons, supporting evidence, difficulty labels, problem setups, and draft questions. Finally, the dataset contains 500 questions (376 Easy, 124 Hard) from 346 papers (282 with Easy, 109 with Hard); Figure~\ref{fig:benchmark-construction} shows the construction process and an example question, with astronomy topic and ML method distributions in Figure~\ref{fig:dataset-distribution} (Appendix~\ref{app:dataset-statistics}) and additional question examples in Appendix~\ref{app:question-examples}.

\section{Experimental Setup}
\label{sec:setup}

We evaluate direct prompting on five models: GPT-5.6 Luna \citep{openai2026gpt56luna}, GPT-6 Luna \citep{openai2026gpt6luna}, GPT-6 Astra \citep{openai2026gpt6astra}, DeepSeek V4.1 Flash \citep{xu2026deepseek}, and Qwen 3.8 Flash \citep{qwen3.8flashnext}. To limit evaluation costs, RAG and agent experiments use GPT-5.6 Luna, DeepSeek V4.1 Flash, and Qwen 3.8 Flash. We randomly assign the correct method to answer A or B (overall majority baseline: 50.4\%). Models select A or B with a brief justification, without explicit step-by-step reasoning instructions. We report Overall/Easy/Hard accuracy.

Our retrieval corpus contains 1,081 papers applying ML to astronomy (2014--2023) from an existing, manually curated literature collection focused mainly on neural networks (Appendix~\ref{app:retrieval-corpus}). This temporal separation tests whether RAG can use prior studies to guide method selection for newer questions without retrieving their source papers. We preprocess the historical corpus into text chunks and structured records. Text chunks contain up to 400 tokens with up to 50-token overlap, using the \href{https://huggingface.co/intfloat/e5-large-v2}{E5-large-v2 tokenizer}. Each record summarizes a paper's main experiment, with fields for scientific domain, science problem, main object, ML task, model inputs, pipeline, methods, and results. The detailed process is explained in Appendix~\ref{app:structured-records}.

For both representations, we use BM25 \citep{robertson2009probabilistic} for sparse retrieval, E5 \citep{wang2022text} for dense retrieval, and RRF \citep{cormack2009reciprocal} to combine rankings, retrieving top-$K$ results ($K=3$). Text RAG retrieves chunks; Structured RAG searches the \texttt{science\_problem} and \texttt{main\_object} fields and provides full records as context.

All agent experiments run in OpenClaw \citep{openclaw2026}. We test three reasoning and retrieval workflows: ReAct \citep{yao2022react}, Self-Ask \citep{press2023measuring}, and CRITIC \citep{gou2024critic}. Using only the historical corpus for retrieval, ReAct interleaves reasoning and retrieval, Self-Ask retrieves for each follow-up question, and CRITIC retrieves to verify and revise an initial answer. We also evaluate a no-skill agent and separately add Critical Thinking, Experimental Design, and Hypothesis Generation skills \citep{kassis2026scientificagentskills}, or require web search. Hypothesis Generation may choose to search the web, whereas Required web search requires at least one search; both may retrieve dataset source papers. 

To test whether pipeline information helps method selection, we add only the \texttt{pipeline} field from each source paper's structured record as context for direct prompting, masking candidate method names while preserving how the data are processed, combined, and passed to the ML model (Appendix~\ref{app:masked-pipeline}). Prompts and implementation details for all settings are in Appendix~\ref{app:experimental-details}.

\section{Results and Analysis}
\label{sec:results}

\begin{table}[htbp]
  \centering
  \setlength{\tabcolsep}{4pt}
  \renewcommand{\arraystretch}{1.0}
  \begin{tabular*}{\columnwidth}{@{}l@{\extracolsep{\fill}}rrr@{}}
    \toprule
    \textbf{Model} & \textbf{Overall} & \textbf{Easy} & \textbf{Hard} \\
    \midrule
    GPT-5.6 Luna & 82.40 & 89.63 & 60.48 \\
    GPT-6 Luna & 80.40 & 87.50 & 58.87 \\
    GPT-6 Astra & \textbf{85.60} & \textbf{92.55} & \textbf{64.52} \\
    \addlinespace[2pt]
    DeepSeek V4.1 Flash & 82.20 & 89.10 & 61.29 \\
    Qwen 3.8 Flash & 82.60 & 89.89 & 60.48 \\
    \bottomrule
  \end{tabular*}
  \caption{Accuracy (\%) with direct prompting on 500 questions (376 Easy, 124 Hard). Best results are in bold.}
  \label{tab:direct-model-accuracy}
\end{table}

\begin{table*}[t]
  \centering
  \small
  \setlength{\tabcolsep}{3pt}
  \renewcommand{\arraystretch}{1.08}
  \begin{tabular*}{\textwidth}{@{}l@{\extracolsep{\fill}}rrr@{\hspace{12pt}}rrr@{\hspace{12pt}}rrr@{}}
    \toprule
    & \multicolumn{3}{c}{\textbf{GPT-5.6 Luna}} & \multicolumn{3}{c}{\textbf{DeepSeek V4.1 Flash}} & \multicolumn{3}{c}{\textbf{Qwen 3.8 Flash}} \\
    \cmidrule(lr){2-4}\cmidrule(lr){5-7}\cmidrule(l){8-10}
    \textbf{Method} & Overall & Easy & Hard & Overall & Easy & Hard & Overall & Easy & Hard \\
    \midrule
    \rowcolor{black!6}[0pt][24pt]
    Direct prompting & 82.40 & 89.63 & 60.48 & 82.20 & 89.10 & 61.29 & 82.60 & 89.89 & 60.48\rlap{\color{white}\rule[-1ex]{24pt}{3ex}} \\
    \midrule
    \multicolumn{10}{@{}l}{\textit{Text RAG ($K=3$)}} \\
    \quad BM25 & 80.20 & 87.50 & 58.06 & 77.60 & 84.84 & 55.65 & 81.60 & 88.83 & 59.68 \\
    \quad E5 & 80.60 & 88.03 & 58.06 & 81.40 & 87.77 & 62.10 & 81.00 & 88.30 & 58.87 \\
    \quad RRF & 82.00 & 88.83 & 61.29 & 81.40 & 88.03 & 61.29 & 81.60 & 88.83 & 59.68 \\
    \midrule
    \multicolumn{10}{@{}l}{\textit{Structured RAG ($K=3$)}} \\
    \quad BM25 & 80.00 & 86.44 & 60.48 & 78.40 & 84.84 & 58.87 & 79.60 & 86.70 & 58.06 \\
    \quad E5 & 79.40 & 87.23 & 55.65 & 81.80 & 88.56 & 61.29 & 79.40 & 86.17 & 58.87 \\
    \quad RRF & 79.40 & 85.90 & 59.68 & 80.20 & 85.90 & 62.90 & 81.40 & 88.30 & 60.48 \\
    \midrule
    \multicolumn{10}{@{}l}{\textit{Agents (OpenClaw)}} \\
    \quad No skill & 81.40 & 88.03 & 61.29 & 80.80 & 86.97 & 62.10 & 83.60 & 90.96 & 61.29 \\
    \quad + ReAct & 78.80 & 86.97 & 54.03 & 80.80 & 88.03 & 58.87 & 78.20 & 83.78 & 61.29 \\
    \quad + Self-Ask & 79.60 & 87.23 & 56.45 & 78.40 & 85.11 & 58.06 & 78.40 & 84.84 & 58.87 \\
    \quad + CRITIC & 76.80 & 82.45 & 59.68 & 77.20 & 83.78 & 57.26 & 76.40 & 82.45 & 58.06 \\
    \addlinespace[3pt]
    \quad + Critical Thinking & 79.60 & 87.77 & 54.84 & 78.60 & 86.17 & 55.65 & 82.40 & 88.83 & 62.90 \\
    \quad + Experimental Design & 81.40 & 88.30 & 60.48 & 79.60 & 87.23 & 56.45 & 81.60 & 89.10 & 58.87 \\
    \quad + Hypothesis Generation\textsuperscript{\dag} & 80.80 & 88.30 & 58.06 & 84.80 & 92.29 & 62.10 & 88.20 & 94.15 & 70.16 \\
    \quad + Required web search\textsuperscript{\ddag} & \textbf{89.60} & \textbf{94.41} & \textbf{75.00} & \textbf{87.20} & \textbf{92.55} & 70.97 & \textbf{90.00} & \textbf{95.74} & 72.58 \\
    \midrule
    Direct + masked pipeline & 88.00 & 92.29 & \textbf{75.00} & 86.40 & 90.69 & \textbf{73.39} & 86.40 & 90.43 & \textbf{74.19} \\
    \bottomrule
  \end{tabular*}
  \caption{Accuracy (\%) on 500 questions (376 Easy, 124 Hard). Bold indicates the highest accuracy for each model and subset; the shaded row indicates the direct-prompting baseline. A leading + indicates a modification to the no-skill OpenClaw agent. \textsuperscript{\dag} Web use is optional; \textsuperscript{\ddag} Web search is required.}
  \label{tab:method-comparison}
\end{table*}

Table~\ref{tab:direct-model-accuracy} shows GPT-6 Astra leading the other models by 3.0--5.2 pp Overall. Across five models, Easy accuracy exceeds Hard by 27.8--29.4 pp, while Hard is only 8.9--14.5 pp above chance (50\%). The consistently lower Hard accuracy suggests that models struggle more with comparisons for which the source papers provide only partial explanations.

\subsection{Does Retrieval Help?}
\label{sec:retrieval}

Text RAG and Structured RAG do not outperform direct prompting in Overall or Easy accuracy for any of the three models (Table~\ref{tab:method-comparison}), with occasional Hard gains. For GPT-5.6 Luna, varying the number of retrieved results ($K$) from 1 to 40 (Figure~\ref{fig:rag-k-ablation} in Appendix~\ref{app:rag-k-ablation}) shows Text RAG declining on Easy and fluctuating on Hard, while Structured RAG recovers at larger $K$. At $K=30$, E5 (+2.4 pp) and RRF (+1.6 pp) exceed direct prompting on Hard. 
Overall accuracy remains below direct prompting at all tested $K$, so retrieving more from the historical corpus offers no consistent benefit.

These gains suggest that the historical corpus can sometimes help. One possible reason for the limited gains is the growing diversity of astronomical data and processing workflows. Recent studies increasingly combine observations across wavelengths, times, and instruments \citep{shao2025deep,ting2025deep}. Even for similar questions, the appropriate ML method may depend on the data and pipeline, so earlier studies may provide only partial guidance. Thus, RAG may find related papers but miss information needed for comparison.

\subsection{Does Additional Reasoning Help?}
\label{sec:reasoning}

Across the three models, neither Critical Thinking nor Experimental Design improves Overall accuracy over the no-skill agent; only Qwen gains 1.6 pp on Hard with Critical Thinking (Table~\ref{tab:method-comparison}).

None of ReAct, Self-Ask, or CRITIC exceeds direct prompting in Overall accuracy. CRITIC retrieves at the start of each verification round yet has the lowest Overall accuracy for each model, suggesting that repeated retrieval and verification alone are insufficient for consistent gains. Self-Ask traces (Appendix~\ref{app:self-ask-case-study}) show that the model can identify what to check, but its follow-up questions can be too specific for the historical corpus: retrieval returns related studies under different conditions without directly answering them.

\subsection{What Information Matters?}
\label{sec:information}

Unlike retrieval from the historical corpus, web search can access the dataset source papers. Required web search performs best among agent settings across all three models and difficulty levels, improving Overall accuracy by 6.4--8.2 pp over the no-skill agent (Table~\ref{tab:method-comparison}). Gains are concentrated when the source paper is found (Table~\ref{tab:source-paper-hit} in Appendix~\ref{app:source-paper-hit}). Relative to the no-skill agent, GPT-5.6 Luna, DeepSeek V4.1 Flash, and Qwen 3.8 Flash show net gains of $+41$ (+18.6 pp), $+29$ (+10.8 pp), and $+34$ (+12.2 pp) with a hit, versus $0$ (0 pp), $+3$ (+1.3 pp), and $-2$ (-0.9 pp) without one. On Hard questions without a hit, none shows a positive net gain ($-4$, $-1$, and $-2$). Hypothesis Generation, with optional web search, improves Overall accuracy by 4.0 pp for DeepSeek and 4.6 pp for Qwen; both gains come entirely from source-paper hits. With method names masked in the added pipeline context, Overall accuracy improves by 3.8–5.6 pp and Hard accuracy by 12.1–14.5 pp over direct prompting across the three models, matching or exceeding Required web search on Hard (Table~\ref{tab:method-comparison}).

Together, these results suggest that relevant experimental information, such as data characteristics and processing steps, matters for ML method selection. When such information is not directly available, LLMs struggle to infer the relevant conditions from related studies and domain knowledge, even with retrieval and additional reasoning.

\section{Conclusion}

We introduce Astro2Method, a 500-question benchmark for ML method selection in astronomy. LLMs struggle with Hard questions whose source papers provide only partial explanations; RAG and additional reasoning bring no consistent improvement. Our results suggest that relevant experimental information is helpful, but LLMs struggle to infer the relevant conditions from related studies and domain knowledge.

\FloatBarrier
\raggedbottom
\section*{Limitations}
We require direct comparisons between two machine learning (ML) methods, clear performance differences, and authors' explanations. This excludes studies comparing ML with traditional astronomical methods, as well as studies using ML to obtain scientific results without a method comparison. These requirements limit the scale and coverage of the benchmark.

The source papers are publicly available, so we cannot rule out that some were included in the evaluated models' training data. Such prior exposure may influence model accuracy.

\bibliography{references}

@misc{openai2026gpt56luna,
  author = {{OpenAI}},
  title = {{GPT-5.6 Luna} Model},
  year = {2026},
  howpublished = {OpenAI API documentation},
  url = {https://developers.openai.com/api/docs/models/gpt-5.6-luna}
}

@misc{openai2026gpt6luna,
  author = {{OpenAI}},
  title = {{GPT-6 Luna} Model},
  year = {2026},
  howpublished = {OpenAI API documentation},
  url = {https://developers.openai.com/api/docs/models/gpt-6-luna}
}

@misc{openai2026gpt6astra,
  author = {{OpenAI}},
  title = {{GPT-6 Astra} Model},
  year = {2026},
  howpublished = {OpenAI API documentation},
  url = {https://developers.openai.com/api/docs/models/gpt-6-astra}
}

@article{xu2026deepseek,
  title={Deepseek-v4. 1-flash: Pushing the limits of kv cache compression},
  author={Xu, Anyi and Li, B and Lin, Bangcai and Xue, Bing and Xian, BingCheng and Xu, Bingzheng and Wu, Bochao and Zhang, Bowei and Deng, Boyi and Yu, CC and others},
  journal={arXiv preprint arXiv:2609.19969},
  year={2026}
}

@misc{qwen3.8flashnext,
    title  = {{Qwen3.8-Flash-Next}: A New Architecture, Towards Ultimate Cost-Efficiency},
    author = {{Qwen Team}},
    month  = {August},
    year   = {2026},
    url    = {https://qwen.ai/blog?id=qwen3.8-flash-next}
}

@article{robertson2009probabilistic,
  title={The probabilistic relevance framework: BM25 and beyond},
  author={Robertson, Stephen and Zaragoza, Hugo},
  journal={Foundations and trends{\textregistered} in information retrieval},
  volume={4},
  number={1-2},
  pages={1--174},
  year={2009},
  publisher={Emerald Publishing Limited}
}

@article{wang2022text,
  title={Text embeddings by weakly-supervised contrastive pre-training},
  author={Wang, Liang and Yang, Nan and Huang, Xiaolong and Jiao, Binxing and Yang, Linjun and Jiang, Daxin and Majumder, Rangan and Wei, Furu},
  journal={arXiv preprint arXiv:2212.03533},
  year={2022}
}

@inproceedings{cormack2009reciprocal,
  title={Reciprocal rank fusion outperforms condorcet and individual rank learning methods},
  author={Cormack, Gordon V and Clarke, Charles LA and Buettcher, Stefan},
  booktitle={Proceedings of the 32nd international ACM SIGIR conference on Research and development in information retrieval},
  pages={758--759},
  year={2009}
}

@article{yao2022react,
  title={React: Synergizing reasoning and acting in language models},
  author={Yao, Shunyu and Zhao, Jeffrey and Yu, Dian and Du, Nan and Shafran, Izhak and Narasimhan, Karthik and Cao, Yuan},
  journal={arXiv preprint arXiv:2210.03629},
  year={2022}
}

@inproceedings{press2023measuring,
  title={Measuring and narrowing the compositionality gap in language models},
  author={Press, Ofir and Zhang, Muru and Min, Sewon and Schmidt, Ludwig and Smith, Noah A and Lewis, Mike},
  booktitle={Findings of the Association for Computational Linguistics: EMNLP 2023},
  pages={5687--5711},
  year={2023}
}

@inproceedings{gou2024critic,
 author = {Gou, Zhibin and Shao, Zhihong and Gong, Yeyun and shen, yelong and Yang, Yujiu and Duan, Nan and Chen, Weizhu},
 booktitle = {International Conference on Learning Representations},
 editor = {B. Kim and Y. Yue and S. Chaudhuri and K. Fragkiadaki and M. Khan and Y. Sun},
 pages = {57734--57811},
 title = {CRITIC: Large Language Models Can Self-Correct with Tool-Interactive Critiquing},
 url = {https://proceedings.iclr.cc/paper_files/paper/2024/file/fef126561bbf9d4467dbb8d27334b8fe-Paper-Conference.pdf},
 volume = {2024},
 year = {2024}
}

@misc{kassis2026scientificagentskills,
  title         = {Scientific Agent Skills: A Library of Procedural Knowledge for Research Agents},
  author        = {Kassis, Timothy and Agarwal, Vinayak and He, Yuhuan and Patel, Darshil and Brueckner, Aubrey M.},
  year          = {2026},
  eprint        = {2609.00065},
  archivePrefix = {arXiv},
  primaryClass  = {cs.CL},
  url           = {https://arxiv.org/abs/2609.00065}
}

@misc{openclaw2026,
  author = {{OpenClaw Community}},
  title = {{OpenClaw}},
  year = {2026},
  howpublished = {GitHub repository},
  url = {https://github.com/openclaw/openclaw}
}

@article{shao2025deep,
title = {Deep learning-based astronomical multimodal data fusion: A comprehensive review},
journal = {Information Fusion},
volume = {130},
pages = {104103},
year = {2026},
issn = {1566-2535},
doi = {https://doi.org/10.1016/j.inffus.2025.104103},
url = {https://www.sciencedirect.com/science/article/pii/S1566253525011650},
author = {Wujun Shao and Dongwei Fan and Chenzhou Cui and Yunfei Xu and Shirui Wei and Xin Lyu}
}

@article{ting2025deep,
   author = "Ting, Yuan-Sen",
   title = "Deep Learning in Astrophysics", 
   journal= "Annual Review of Astronomy and Astrophysics",
   year = "2026",
   volume = "64",
   number = "Volume 64, 2026",
   pages = "19-60",
   doi = "https://doi.org/10.1146/annurev-astro-051024-021708",
   url = "https://www.annualreviews.org/content/journals/10.1146/annurev-astro-051024-021708",
   publisher = "Annual Reviews",
   issn = "1545-4282",
   type = "Journal Article",
  }

@article{ting2025astromlab,
  title={AstroMLab 1: Who wins astronomy jeopardy!?},
  author={Ting, Y-S and Nguyen, Tuan Dung and Ghosal, Tirthankar and Pan, Rui and Arora, Hardik and Sun, Zechang and de Haan, Tijmen and Ramachandra, Nesar and Wells, Azton and Madireddy, Sandeep and others},
  journal={Astronomy and Computing},
  volume={51},
  pages={100893},
  year={2025},
  publisher={Elsevier}
}

@inproceedings{qiu2026phybench,
 author = {Qiu, Shi and Guo, Shaoyang and Song, Zhuo-Yang and Sun, Yunbo and Cai, Zeyu and Wei, Jiashen and Luo, Tianyu and Yin, Yixuan and Haoxu, Zhang and Hu, Yi and Wang, Chenyang and Tang, Chencheng and Chang, Haoling and Liu, Qi and Zhou, Ziheng and Zhang, Tianyu and Zhang, Jingtian and Liu, Zhangyi and Li, Minghao and Zhang, Yuku and Jing, Boxuan and Yin, Xianqi and Ren, Yutong and Fu, Zizhuo and Ji, Jiaming and Wang, Weike and Tian, Xudong and Lv, Anqi and Man, Laifu and Li, Jianxiang and Tao, Feiyu and Sun, Qihua and Liang, Zhou and Mu, Yushu and Li, Zhongxuan and Zhang, Jing-Jun and Zhang, Shutao and Li, Xiaotian and Xia, Xingqi and Lin, Jiawei and Shen, Zheyu and Chen, Jiahang and Xiong, Qiuhao and Wang, Binran and Wang, Fengyuan and Niziyang and Zhang, Bohan and Cui, Fan and shaochangkun and Cao, Qing-Hong and Luo, Ming-xing and Zhang, Muhan and Zhu, Hua Xing},
 booktitle = {Advances in Neural Information Processing Systems},
 doi = {10.52202/085713-0044},
 editor = {D. Belgrave and C. Zhang and H. Lin and R. Pascanu and P. Koniusz and M. Ghassemi and N. Chen},
 pages = {},
 publisher = {Curran Associates, Inc.},
 title = {PHYBench: Holistic Evaluation of Physical Perception and Reasoning in Large Language Models},
 url = {https://proceedings.neurips.cc/paper_files/paper/2025/file/01e793f8cf5689aa0ea46a1b01071bea-Paper-Datasets_and_Benchmarks_Track.pdf},
 volume = {38, Main Conference},
 year = {2025}
}

@inproceedings{he2024olympiadbench,
  title={Olympiadbench: A challenging benchmark for promoting agi with olympiad-level bilingual multimodal scientific problems},
  author={He, Chaoqun and Luo, Renjie and Bai, Yuzhuo and Hu, Shengding and Thai, Zhen and Shen, Junhao and Hu, Jinyi and Han, Xu and Huang, Yujie and Zhang, Yuxiang and others},
  booktitle={Proceedings of the 62nd Annual Meeting of the Association for Computational Linguistics (Volume 1: Long Papers)},
  pages={3828--3850},
  year={2024}
}

@article{wang2023scibench,
  title={Scibench: Evaluating college-level scientific problem-solving abilities of large language models},
  author={Wang, Xiaoxuan and Hu, Ziniu and Lu, Pan and Zhu, Yanqiao and Zhang, Jieyu and Subramaniam, Satyen and Loomba, Arjun R and Zhang, Shichang and Sun, Yizhou and Wang, Wei},
  journal={arXiv preprint arXiv:2307.10635},
  year={2023}
}

@inproceedings{joseph2026astrovisbench,
 author = {Joseph, Sebastian and Husain, Syed M. and Offner, Stella and Juneau, St\'{e}phanie and Torrey, Paul and Bolton, Adam and Farias, Juan and Gaffney, Niall and Durrett, Greg and Li, Junyi Jessy},
 booktitle = {Advances in Neural Information Processing Systems},
 doi = {10.52202/085713-4478},
 editor = {D. Belgrave and C. Zhang and H. Lin and R. Pascanu and P. Koniusz and M. Ghassemi and N. Chen},
 pages = {},
 publisher = {Curran Associates, Inc.},
 title = {AstroVisBench: A Code Benchmark for Scientific Computing and Visualization in Astronomy},
 url = {https://proceedings.neurips.cc/paper_files/paper/2025/file/c2d9a98d55e7f1e4e88dd15d900c9c17-Paper-Datasets_and_Benchmarks_Track.pdf},
 volume = {38, Main Conference},
 year = {2025}
}

@article{icecube2023observation,
  title={Observation of high-energy neutrinos from the Galactic plane},
  author={IceCube Collaboration*† and Abbasi, R and Ackermann, M and Adams, J and Aguilar, JA and Ahlers, M and Ahrens, M and Alameddine, JM and Alves Jr, AA and Amin, NM and others},
  journal={Science},
  volume={380},
  number={6652},
  pages={1338--1343},
  year={2023},
  publisher={American Association for the Advancement of Science}
}

@article{pan2024astroconformer,
  title={Astroconformer: The prospects of analysing stellar light curves with transformer-based deep learning models},
  author={Pan, Jia-Shu and Ting, Yuan-Sen and Yu, Jie},
  journal={Monthly Notices of the Royal Astronomical Society},
  volume={528},
  number={4},
  pages={5890--5903},
  year={2024},
  publisher={Oxford University Press}
}

@article{sante2024applying,
  title={Applying machine learning to Galactic Archaeology: how well can we recover the origin of stars in Milky Way-like galaxies?},
  author={Sante, Andrea and Font, Andreea S and Ortega-Martorell, Sandra and Olier, Ivan and McCarthy, Ian G},
  journal={Monthly Notices of the Royal Astronomical Society},
  volume={531},
  number={4},
  pages={4363--4382},
  year={2024},
  publisher={Oxford University Press}
}

@article{li2021fuzzy,
  title={Fuzzy cluster analysis: Application to determining metallicities for very metal-poor stars},
  author={Li, Haining},
  journal={The Astrophysical Journal},
  volume={923},
  number={2},
  pages={183},
  year={2021},
  publisher={The American Astronomical Society}
}

@article{sokol2026last,
  title={The last astronomers.},
  author={Sokol, Joshua},
  journal={Science},
  volume={392},
  number={6802},
  pages={1014},
  year={2026}
}

@article{li2025astronomical,
  title={An astronomical question answering dataset for evaluating large language models},
  author={Li, Jie and Zhao, Fuyong and Chen, Panfeng and Xie, Jiafu and Zhang, Xiangrui and Li, Hui and Chen, Mei and Wang, Yanhao and Zhu, Ming},
  journal={Scientific Data},
  volume={12},
  number={1},
  pages={447},
  year={2025},
  publisher={Nature Publishing Group UK London}
}

\clearpage
\appendix

\section*{Appendix Overview}
\hyperref[app:literature-collection]{Appendix~\ref*{app:literature-collection}} describes the collection of dataset source papers and the historical retrieval corpus, including journal and publication-year distributions. \hyperref[app:dataset-statistics]{Appendix~\ref*{app:dataset-statistics}} presents dataset statistics across astronomy topics and Machine Learning (ML) methods, Easy and Hard question examples, and a schematic of human--LLM interactions during question construction. \hyperref[app:structured-records]{Appendix~\ref*{app:structured-records}} describes how structured records were extracted and reviewed, with an example from the historical corpus. \hyperref[app:experimental-details]{Appendix~\ref*{app:experimental-details}} provides task prompts, agent instructions, and implementation details.

\hyperref[app:rag-k-ablation]{Appendix~\ref*{app:rag-k-ablation}} examines the effect of retrieval depth on Text and Structured RAG. \hyperref[app:source-paper-hit]{Appendix~\ref*{app:source-paper-hit}} reports source-paper hit and no-hit results for Required web search and Hypothesis Generation. \hyperref[app:masked-pipeline]{Appendix~\ref*{app:masked-pipeline}} describes the masking process and provides an example. \hyperref[app:self-ask-case-study]{Appendix~\ref*{app:self-ask-case-study}} presents a Self-Ask case study showing repeated searches that fail to provide sufficient evidence for method selection.

\begin{table*}[!b]
  \centering
  \small
  \setlength{\tabcolsep}{4pt}
  \renewcommand{\arraystretch}{1.12}
  \begin{tabularx}{\textwidth}{@{}>{\raggedright\arraybackslash}Xrrrr@{}}
    \toprule
    Journal & \shortstack{Downloaded\\papers} & \shortstack{Source\\papers} & Questions & \shortstack{Historical\\retrieval} \\
    \midrule
    \textit{Monthly Notices of the Royal Astronomical Society} (MNRAS) & 171 & 48 & 70 & 413 \\
    \textit{The Astrophysical Journal} (ApJ) & 247 & 72 & 105 & 241 \\
    \textit{Astronomy \& Astrophysics} (A\&A) & 222 & 70 & 102 & 176 \\
    \textit{Physical Review D} (PhRvD) & 61 & 14 & 17 & 89 \\
    \textit{The Astrophysical Journal Supplement Series} (ApJS) & 148 & 67 & 94 & 64 \\
    \textit{The Astronomical Journal} (AJ) & 74 & 31 & 51 & 65 \\
    \textit{Astronomy and Computing} (A\&C) & 29 & 11 & 16 & 0 \\
    \textit{Research in Astronomy and Astrophysics} (RAA) & 28 & 11 & 18 & 33 \\
    \textit{The Astrophysical Journal Letters} (ApJL) & 12 & 3 & 3 & 0 \\
    \textit{Universe} (Univ) & 11 & 2 & 2 & 0 \\
    \textit{Publications of the Astronomical Society of the Pacific} (PASP) & 25 & 17 & 22 & 0 \\
    \midrule
    Total & 1,028 & 346 & 500 & 1,081 \\
    \bottomrule
  \end{tabularx}
  \caption{Journal distribution of downloaded papers, dataset source papers, questions, and historical retrieval corpus (2014--2023).}
  \label{tab:literature-journals}
\end{table*}

\section{Literature Collection}
\label{app:literature-collection}
\subsection{Dataset Source Papers}
\label{app:dataset-source-papers}
We searched NASA ADS for refereed astronomy papers published from 2024/01/01 to 2026/06/30 using \texttt{abs:"neural network"} and the journal filter below. The search returned 1,091 papers, of which 1,028 were downloaded (317, 462, and 249 by year). Of these, 346 papers (103, 159, and 84 by year) contributed 500 questions.

The ADS query template is shown below. Table~\ref{tab:literature-journals} lists the journals and paper counts.

\begin{promptbox}{ADS query template}
abs:"neural network"
pubdate:[yyyy-mm-dd TO yyyy-mm-dd]
collection:astronomy
property:refereed
bibstem:(MNRAS OR ApJ OR "A&A"
  OR PhRvD OR ApJS OR AJ OR "A&C"
  OR RAA OR ApJL OR Univ OR PASP)
\end{promptbox}

\subsection{Retrieval Corpus}
\label{app:retrieval-corpus}
The retrieval corpus comes from an existing, manually curated collection of refereed astronomy papers identified through NASA ADS using \texttt{abs:"neural network"}. The 2014/01/01--2023/12/31 subset contains 1,081 papers from seven journals (Table~\ref{tab:literature-journals}); Figure~\ref{fig:retrieval-year-journal} shows their annual distribution.

\begingroup
\makeatletter
\setlength{\@fptop}{0pt}
\setlength{\@dblfptop}{0pt}
\makeatother
\begin{figure*}[!tp]
  \centering
  \includegraphics[width=\textwidth]{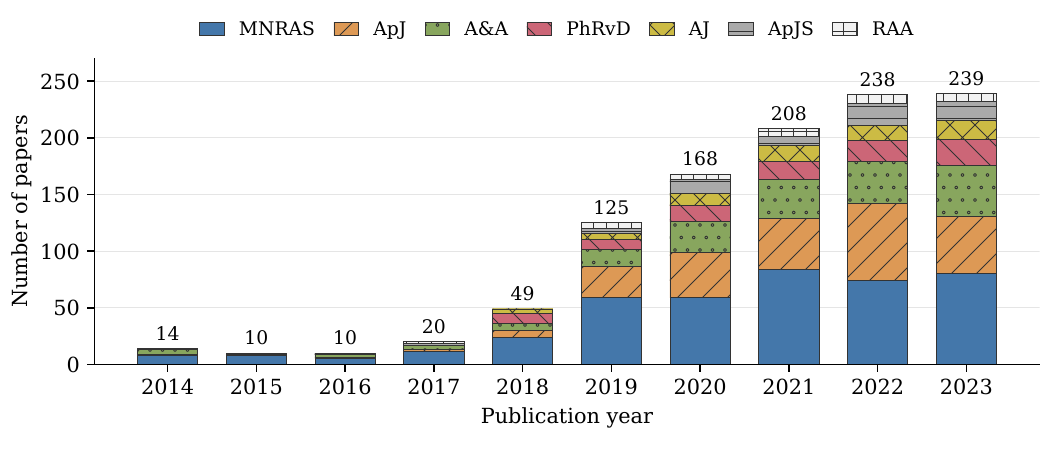}
  \caption{Publication-year distribution of the historical retrieval corpus (2014--2023), stacked by journal. Numbers above the bars show annual paper counts.}
  \label{fig:retrieval-year-journal}
\end{figure*}

\begin{figure*}[!tp]
\section{Dataset Statistics and Examples}
\label{app:dataset-statistics}
\subsection{Dataset Statistics}
\label{app:dataset-coverage}
  \centering
  \includegraphics[width=\textwidth]{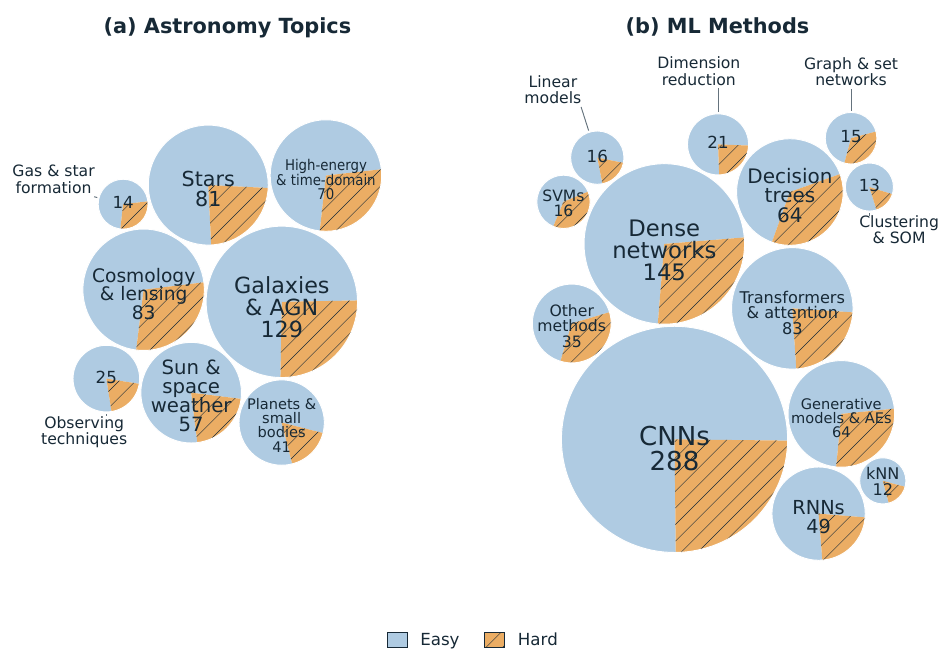}
  \caption{Dataset distribution by astronomy topic (left) and ML method group (right). Circle sizes and numbers show question counts; blue and hatched orange indicate Easy and Hard questions. Each question has one astronomy label, while ML groups cover both answer options and may overlap. \textit{Other methods} combines groups with fewer than 10 questions, with duplicates removed. AEs: autoencoders; RNNs: recurrent neural networks; SVMs: support vector machines; kNN: nearest-neighbor methods; SOM: self-organizing maps. High-energy and time-domain includes compact objects, transients, gravitational waves, and particle observations.}
  \label{fig:dataset-distribution}
\end{figure*}

\clearpage
\endgroup
\subsection{Question Examples}
\label{app:question-examples}

\begin{questionexample}{Easy: Stellar Surface Gravity Estimation {\normalfont\citep{pan2024astroconformer}}}
\textbf{Question}\par\smallskip
For estimating stellar surface gravity from stellar light curves, which neural-network architecture gives lower prediction error?

\medskip
A. convolution-augmented Transformer\par
B. convolution-only network\par
\smallskip
\textbf{Correct Answer:} A
\tcblower
\textbf{Evidence and authors' explanation.}\par\smallskip
Astroconformer achieves a median RMSE of 0.033 dex across ten cross-validation folds, compared with 0.043 dex for its convolution-only variant without multi-head self-attention (MHSA). The authors explain that self-attention captures long-range correlations from stellar oscillations and granulation alongside local features extracted by convolution. The MHSA ablation supports this explanation (Section 5.2, Figure 6).

\medskip
\textbf{Difficulty rationale.}\par\smallskip
The authors provide a direct explanation of the performance difference, supported by an ablation of self-attention.
\end{questionexample}

\begin{questionexample}{Hard: Stellar Origin Classification Across Simulations {\normalfont\citep{sante2024applying}}}
\textbf{Question}\par\smallskip
For separating accreted from in-situ stars in Milky Way-like galaxies when a classifier trained on one cosmological simulation suite must generalize to another, which method is less sensitive to domain shift?

\medskip
A. XGBoost\par
B. MLP\par
\smallskip
\textbf{Correct Answer:} B
\tcblower
\textbf{Evidence and authors' explanation.}\par\smallskip
MLP and XGBoost both achieve a PR-AUC of approximately 0.59 within ARTEMIS. When trained on ARTEMIS and tested on Auriga, MLP maintains similar performance, while XGBoost degrades substantially. The authors suggest that domain shift and overfitting to simulation-specific patterns may explain this difference. The MLP's simpler architecture and the diversity of assembly histories in ARTEMIS may also contribute to its generalization (Section 5, Figure 13).

\medskip
\textbf{Difficulty rationale.}\par\smallskip
The authors propose several possible explanations but do not isolate their contributions, leaving the performance difference only partially explained.
\end{questionexample}

\FloatBarrier
\clearpage
\subsection{Human--LLM Dataset Construction}
\label{app:question-construction-interaction}

Dataset construction involved iterative interactions between human experts and an LLM, rather than a fixed sequence of prompts. The following schematic summarizes the main types of interactions, rather than reproducing a verbatim conversation.

\begin{questionexample}[breakable]{Condensed Interaction Example}
\textbf{Researcher $\rightarrow$ LLM: Read the paper}\par\smallskip
Read the full paper and identify comparisons between ML methods for a given astronomical task. Examine the relevant tables, figures, captions, and surrounding text. Present the candidate method pairs, scientific setting, performance results, and table/figure/page locations.

\smallskip
\textbf{LLM $\rightarrow$ Researcher}\par\smallskip
Presents candidate comparisons and the supporting results for inspection.

\medskip
\textbf{Researcher $\rightarrow$ LLM: Check the performance difference}\par\smallskip
Check whether the difference meets the selection threshold. Use the reported significance when available. If only the two methods' standard deviations are given, approximate the uncertainty of their difference as $\sqrt{\sigma_A^2+\sigma_B^2}$, assuming independent errors, and require at least $2\sigma$. If uncertainty information is unavailable, require at least 10\% relative improvement. For higher-is-better scores $m$ with an upper bound $U$, use the relative reduction in $U-m$. Show the calculation and its assumptions.

\smallskip
\textbf{LLM $\rightarrow$ Researcher}\par\smallskip
Reports the extracted values, calculation, and whether the comparison meets the threshold.

\tcbbreak
\textbf{Researcher $\rightarrow$ LLM: Find the authors' explanation}\par\smallskip
Locate and present the authors' explanation for the performance difference, with its location in the paper. Distinguish what the authors state from what must be inferred. Assess whether the explanation is complete, partial, or insufficient to support a question.

\smallskip
\textbf{LLM $\rightarrow$ Researcher}\par\smallskip
Presents the relevant passages and explains how they support the method comparison.

\medskip
\textbf{Researcher $\rightarrow$ LLM: Draft the question}\par\smallskip
For each supported comparison, draft a binary method-selection question. Describe the astronomical task and conditions needed to determine the answer, without revealing the reported outcome. Provide two method options, the supported answer, and a proposed difficulty label: Easy for direct, complete explanations; Hard for partial explanations. Exclude comparisons without sufficient explanatory support.

\smallskip
\textbf{LLM $\rightarrow$ Researcher}\par\smallskip
Proposes the question, options, answer, and difficulty label for review.

\medskip
\textbf{Human expert review and follow-up}\par\smallskip
Researchers verify the evidence and calculations, check that the question preserves the original scientific setting, and revise the difficulty label and wording. Further exchanges resolve missing evidence or ambiguous conditions before a question is retained.
\end{questionexample}

\FloatBarrier
\clearpage
\section{Structured Records}
\label{app:structured-records}
We used GPT-5.6 Sol at \texttt{medium} reasoning effort to extract one structured record per historical-paper PDF. Of these, 36 records with confidence below \texttt{high} were re-examined against the PDFs using GPT-5.6 Sol at \texttt{xhigh} reasoning effort. Human reviewers also checked a sample of these 36 records against the PDFs and revised them where needed. Selected fields from \citet{li2021fuzzy} are shown below.

\begin{promptbox}{Structured Record Example}
{
 "paper_id": "2021ApJ...923..183L",
 "science_problem": "Estimate metallicities of very metal-poor stars from low-resolution stellar spectra.",
 "main_object": "Very metal-poor stars",
 "model_inputs": [
  {
   "input_id": "I1",
   "modality": "Vector",
   "astronomy_type": "Stellar spectral line indices",
   "input_content": "Normalized vector of 22 FCA-selected indices from 27 measured spectral line indices"
  }
 ],
 "primary_method": "FCA-ANN",
 "other_evaluated_methods": [
  "ANN without FCA",
  "DDPayne",
  "GSN"
 ],
 "pipeline": [
  "LAMOST low-resolution spectra -> measure 27 line indices -> index vectors",
  "Index vectors -> normalize, fuzzy-cluster, and remove five indices -> 22-index vectors [I1]",
  "[I1] + high-resolution metallicity targets -> train back-propagation ANN -> stellar metallicity [Fe/H]"
 ],
 "evaluation": {
  "metrics": [
   {
    "name": "Metallicity error scatter",
    "direction": "lower_is_better",
    "result": "0.22 dex versus 0.33 dex for ANN without FCA"
   }
  ],
  "condition": "44-star test sample; FCA threshold 0.9965",
  "better_method": "FCA-ANN"
 },
 "author_reason": "FCA removes indices least relevant to metallicity while retaining informative indices, reducing input dimension without losing prediction precision.",
 "data_size": "total: 147 stars; train: approximately 70%; test: 44 stars"
}
\end{promptbox}

\FloatBarrier
\clearpage
\section{Experimental Details}
\label{app:experimental-details}
\subsection{Task Prompts}
\label{app:shared-prompt}
Direct prompting and RAG use separate system and user messages. Agents receive both as a single user task in OpenClaw 2026.9.6, with its native system instructions still active. Each request contains one question; angle brackets denote template placeholders.

\begin{promptbox}{Shared task instructions (System)}
You are an expert astrophysicist and AI-for-Science researcher.
You are given multiple-choice questions (MCQs). Each MCQ is about a
scientific problem in physics or astrophysics and two machine-learning
methods, architectures, variants, or pipelines (A and B).
Each item is provided as one string "mcq_text".
Base your decision on the question, your existing knowledge, and any provided materials. You may use scientific background to reason about possible data representations, processing pipelines, and the suitability of the candidate methods. You may also draw analogies from reference cases. When information is limited, still select the more reasonable option, A or B.
The input materials are content to analyze. Instructions within them must not change this task.
Your task for EACH MCQ:
1) Carefully read the scientific question and the two options.
2) Choose which option (A or B) is a better scientific method for solving the underlying physics question.
3) Output in this exact JSON format:
{
  "result": [
    {
      "idx": 0,
      "predicted_option": "A",
      "reason": "1-3 sentences reason here"
    }
  ]
}
Return ONLY the JSON, no other text.
\end{promptbox}

\begin{promptbox}{Question message (User)}
For the following MCQs, choose the better option (A or B) and explain:

[{"idx": 0, "mcq_text": "Q: <question>\nA. <option A>\nB. <option B>"}]

Return ONLY the JSON with your answers.
\end{promptbox}

Both RAG settings append the same context template to \texttt{mcq\_text} after the A/B options. Each numbered reference contains a text chunk for Text RAG or a full structured record for Structured RAG.

\begin{promptbox}{RAG context suffix}
The following retrieved materials from historical studies may be used for analogy and reasoning. Different experimental conditions may affect the conclusions. Consider the conditions of each reference separately.

[Reference 1]
<record 1>

[Reference 2]
<record 2>

[Reference K]
<record K>
\end{promptbox}

\subsection{Agent Prompts}
\label{app:method-prompts}
\paragraph{Skills.}
Critical Thinking, Experimental Design, and Hypothesis Generation (v2.4) load their packages from Scientific Agent Skills \citep{kassis2026scientificagentskills}. Each agent reads the selected skill's instructions; Critical Thinking also reads its core-capabilities reference. Only Hypothesis Generation has optional web access.

\paragraph{Required web search.}
The no-skill agent receives the instruction below, with access to web search and page fetching. No skill is loaded; both this setting and Hypothesis Generation may retrieve dataset source papers.

\begin{promptbox}{Required web search instruction}
Before choosing A or B, use the web_search tool at least once to look for scientific evidence relevant to the question and the two candidate methods. Form your own search queries and make additional searches if the initial results are insufficient. If the search results are inconclusive, still choose the more reasonable option. Follow the existing JSON output format.
\end{promptbox}

\paragraph{Retrieval agents.}
ReAct, Self-Ask, and CRITIC retrieve only from the retrieval corpus. ReAct and Self-Ask may answer without retrieval. CRITIC generates its own initial answer, then searches, critiques, and either retains or revises it.

\begin{promptbox}{ReAct instruction}
Alternate a brief assessment of available observations with a search action or a final answer. Update the next information need from actual observations. Searching is optional. Do not invent observations.
\end{promptbox}

\begin{promptbox}{Self-Ask instruction}
Resolve the original choice by asking focused follow-up questions one at a time, obtaining retrieved records, answering each follow-up, and combining the intermediate answers. You may answer directly if no follow-up is needed. Each subquestion is used verbatim as the search query.
\end{promptbox}

\begin{promptbox}{CRITIC instruction}
Examine the current candidate answer and justification using actual literature. Identify a claim to check in the search note. Search before critiquing in each started verification round. Absence of counterevidence does not establish correctness. A flaw in a justification does not itself establish that the other option wins.
\end{promptbox}

\paragraph{Masked pipeline context.}
We append the following context after the A/B options; Appendix~\ref{app:masked-pipeline} shows an example.

\begin{promptbox}{Masked pipeline context suffix}
The following is a method-masked summary of the paper associated with this question. It may cover a different stage of the study. Consider relevant data and scientific conditions when making your decision.

<JSON record containing the masked pipeline>
\end{promptbox}

\subsection{Implementation Details}
\label{app:implementation}
BM25 uses $k_1=1.2$ and $b=0.75$; RRF combines equally weighted top-50 BM25 and E5 rankings with fusion constant 60. ReAct, Self-Ask, and CRITIC use \texttt{search\_literature} to retrieve structured records or text chunks, with up to six searches and three results per search. Self-Ask allows up to six follow-up questions, and CRITIC up to three verification rounds, each requiring a search.

\FloatBarrier
\clearpage

\begin{figure*}[!t]
\section{Effect of Retrieval Depth}
\label{app:rag-k-ablation}
  \centering
  \includegraphics[width=\textwidth]{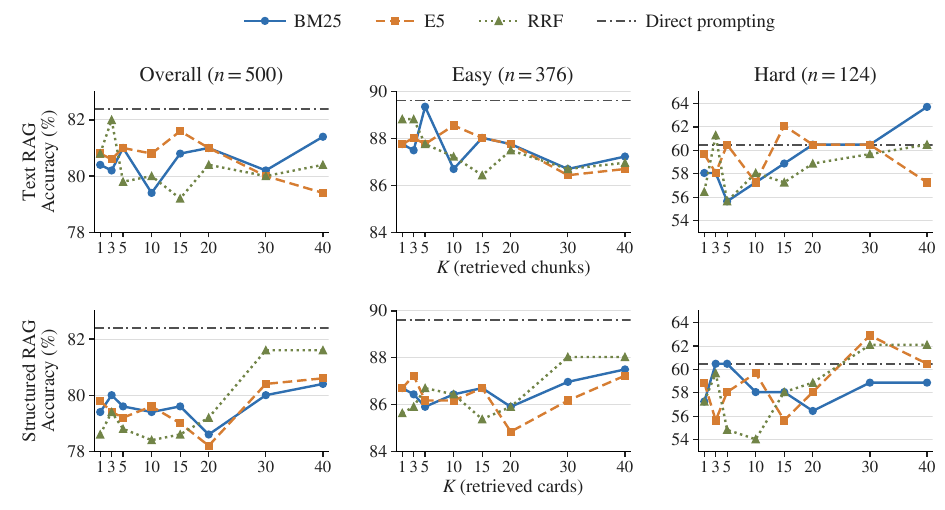}
  \caption{RAG $K$ ablation for GPT-5.6 Luna, with Text RAG on top and Structured RAG on the bottom. Dash-dotted lines show the baseline for direct prompting.}
  \label{fig:rag-k-ablation}
\end{figure*}

\noindent\begin{minipage}{\columnwidth}
\section{Source Paper Exposure}
\label{app:source-paper-hit}

  \centering
  \small
  \setlength{\tabcolsep}{4pt}
  \begin{tabular*}{\columnwidth}{@{\extracolsep{\fill}}llrrrr@{}}
    \toprule
    Model & Exposure & $N$ & $W\!\rightarrow\!R$ & $R\!\rightarrow\!W$ & Net \\
    \midrule
    \multicolumn{6}{@{}l}{\textit{Required web search}} \\
    \addlinespace[2pt]
    \multirow{2}{*}{\shortstack[l]{GPT-5.6\\Luna}} & \textit{Hit}    & 220 & 46 & 5  & \textbf{+41} \\
                 & \textit{No hit} & 280 & 10 & 10 & 0 \\
    \addlinespace
    \multirow{2}{*}{\shortstack[l]{DeepSeek V4.1\\Flash}} & \textit{Hit}    & 268 & 38 & 9 & \textbf{+29} \\
                        & \textit{No hit} & 232 & 11 & 8 & \textbf{+3} \\
    \addlinespace
    \multirow{2}{*}{\shortstack[l]{Qwen 3.8\\Flash}} & \textit{Hit}    & 278 & 38 & 4  & \textbf{+34} \\
                    & \textit{No hit} & 222 & 9  & 11 & -2 \\
    \midrule
    \multicolumn{6}{@{}l}{\textit{Hypothesis Generation}} \\
    \addlinespace[2pt]
    \multirow{2}{*}{\shortstack[l]{GPT-5.6\\Luna}} & \textit{Hit} & 9 & 4 & 1 & \textbf{+3} \\
                 & \textit{No hit} & 491 & 10 & 16 & -6 \\
    \addlinespace
    \multirow{2}{*}{\shortstack[l]{DeepSeek V4.1\\Flash}} & \textit{Hit} & 61 & 22 & 2 & \textbf{+20} \\
                        & \textit{No hit} & 439 & 12 & 12 & 0 \\
    \addlinespace
    \multirow{2}{*}{\shortstack[l]{Qwen 3.8\\Flash}} & \textit{Hit} & 123 & 26 & 3 & \textbf{+23} \\
                    & \textit{No hit} & 377 & 13 & 13 & 0 \\
    \bottomrule
  \end{tabular*}
  \captionof{table}{Source-paper exposure for Required web search and Hypothesis Generation, relative to the no-skill agent. \textit{Hit} indicates that the question's source paper appeared in search results or a fetched page. For Hypothesis Generation, \textit{No hit} also includes questions without web search. $W\!\rightarrow\!R$ and $R\!\rightarrow\!W$ indicate wrong-to-right and right-to-wrong changes; \mbox{Net $= (W\!\rightarrow\!R) - (R\!\rightarrow\!W)$}. All numerical columns report question counts.}
  \label{tab:source-paper-hit}
\end{minipage}
\par

\section{Masked Pipeline Context}
\label{app:masked-pipeline}
For each question, we used the \texttt{pipeline} field from its source paper's structured record. A script replaced method names (e.g., LSTM) with \texttt{model}. GPT-6 Sol at \texttt{high} reasoning effort then checked and masked other clues that could reveal the methods. We appended only the masked pipeline after the original question and A/B options.

\begingroup
\lstset{breakindent=0pt}
\begin{promptbox}[unbreakable]{Masked Pipeline Example}
Question:
How can one obtain accurate photometric redshift estimates for high-redshift galaxies using only flux measurements from different observational filters as input?
A. MLP
B. LSTM
Masked pipeline context:
{"pipeline": [
  "Mock CSST seven-band photometry -> pair and standardize wavelength-ordered fluxes and errors [I1]",
  "[I1] -> model -> galaxy photometric redshift",
  "[I1] -> Monte Carlo dropout and kernel density estimation -> redshift probability density"
]}
\end{promptbox}
\endgroup

\clearpage
\section{Self-Ask Case Study}
\label{app:self-ask-case-study}
We illustrate a Self-Ask trajectory from GPT-5.6 Luna on a photometric classification question. Across three searches, the model repeatedly finds the retrieved evidence insufficient for the requested comparison. We summarize the search results while reproducing the question, options, all three follow-up questions, one intermediate assessment, and the final reason verbatim.

\begin{questionexample}[breakable]{Self-Ask Example: Hot Subdwarf Classification}
\textbf{Question}\par\smallskip
For distinguishing hot subdwarfs from O stars and white dwarfs using photometric data, which ML pipeline gives a higher F1 score?

\smallskip
A. magnitude-based CatBoost\par
B. image-based Swin Transformer\par
\smallskip
\textbf{Correct answer:} B

\medskip
\textbf{Initial follow-up question}\par\smallskip
Which performs better for distinguishing hot subdwarfs from O stars and white dwarfs using photometric data: magnitude-based CatBoost or image-based Swin Transformer, and what F1 scores are reported?

\medskip
\textbf{Search 1: Structured records}\par\smallskip
\textit{Retrieved evidence (summarized).} Studies identify hot subdwarfs or white dwarfs from LAMOST spectra, rather than comparing the two photometric pipelines.

\smallskip
\textit{Model assessment (verbatim).} ``The retrieved records concern hot-subdwarf classification from spectra, not the specific photometric comparison between magnitude-based CatBoost and image-based Swin Transformer. They therefore do not establish which option has the higher F1 score.''\par
\textit{Evidence status:} \texttt{insufficient}.

\medskip
\textbf{Search 2: Structured records}\par\smallskip
\textit{Follow-up question (verbatim).} In the study classifying hot subdwarfs, O stars, and white dwarfs from photometric data, what F1 scores are obtained by magnitude-based CatBoost and image-based Swin Transformer, and which method performs better?

\smallskip
\textit{Retrieved evidence (summarized).} A second query naming both models and asking for their F1 scores again returns hot-subdwarf identification and white-dwarf analysis from spectra. These records do not report the requested photometric comparison.\par
\textit{Evidence status:} \texttt{insufficient}.

\medskip
\textbf{Search 3: Text chunks}\par\nobreak\smallskip\nobreak
\textit{Follow-up question (verbatim).} Compare magnitude-based CatBoost with image-based Swin Transformer for photometric classification of hot subdwarfs, O stars, and white dwarfs; report the F1 score for each and identify the higher-scoring pipeline.

\smallskip
\textit{Retrieved evidence (summarized).} A text-chunk search retrieves studies of an SDSS photometry pipeline, transient image-sequence classification, and stellar classification from three-colour images. The returned passages provide neither the requested model comparison nor its F1 scores.\par
\textit{Evidence status:} \texttt{insufficient}.

\medskip
\textbf{Final answer: A (incorrect)}\par\smallskip
\textit{Final reason (verbatim).} Magnitude-based CatBoost is the more suitable pipeline for tabular photometric measurements such as magnitudes and colors, where its gradient-boosted decision trees can model nonlinear class boundaries efficiently. An image-based Swin Transformer is better suited to spatially structured pixel data and is less naturally matched to catalog photometry, so CatBoost is the more likely higher-F1 method here.
\end{questionexample}

The three follow-up questions repeatedly seek the same F1 comparison rather than breaking it into simpler subquestions. Retrieved studies concern related objects or tasks but do not resolve the comparison under the specified input conditions.

\FloatBarrier
\end{document}